\documentclass{jpsj3}
\usepackage{txfonts}
\usepackage{color}

\title{Charge Oscillations Emerging after Application of an Intense Light Field to Superconductors on a Dimer Lattice}

\author{Kenji Yonemitsu\thanks{E-mail: kxy@phys.chuo-u.ac.jp}}
\inst{Department of Physics, Chuo University, Bunkyo, Tokyo 112-8551, Japan} 

\abst{ Motivated by recent experimental findings for an organic superconductor, charge oscillations that emerge after a strong pulse of an oscillating electric field is applied are studied in electron systems in a superconducting phase on a lattice with a dimerized structure. They are analyzed using Fourier spectra of charge densities whose time profiles are obtained by numerically solving the time-dependent Schr\"odinger equation within a mean-field approximation. Depending on the strengths of attractive interactions, different charge-oscillation modes appear. For weak attractions, the charge-oscillation mode is an electronic breathing mode, which was previously found for repulsive interactions. For strong attractions, it is a pair analog of the electronic breathing mode (a ``pair'' breathing mode). For intermediate attractions, it is another mode whose transient current distributions are considerably different from those of the breathing modes. We investigate how their frequencies and amplitudes depend on interactions and transfer integrals. }

\begin{document}
\maketitle

\section{Introduction}
In solid-state materials, various cooperative and dynamic phenomena are known. Photoinduced phase transitions are such phenomena, and they are achieved between different phases and on different timescales.\cite{koshigono_jpsj06,yonemitsu_pr08,basov_rmp11,nicoletti_aop16,giannetti_aip16} In many cases, electronic orders are melted or weakened. However, in some cases, orders are transiently constructed or enhanced.\cite{fausti_s11,ishikawa_ncomms14,mitrano_n16,rettig_ncomms16,singer_prl16,mor_prl17} 

As a possible mechanism for the enhancement of an electronic order, dynamical localization may work,\cite{dunlap_prb86,grossmann_prl91,kayanuma_pra94}  which suppresses the itinerancy of electrons and enhances the relative importance of interactions during and even after photoexcitation.\cite{tsuji_prl11,tsuji_prb12,nishioka_jpsj14,yonemitsu_jpsj15,yanagiya_jpsj15,ono_prb16,ono_prb17} In some cases, however, it has been shown that dynamical localization is insufficient and electron correlations are necessary to explain experimentally observed details in the enhancement.\cite{ishikawa_ncomms14,kawakami_prb17,yonemitsu_jpsj17a,yonemitsu_jpsj17b} For excitonic insulators,\cite{mor_prl17} the excitonic order has been theoretically shown to be enhanced by different mechanisms.\cite{murakami_prl17,tanaka_prb18,tanabe_arx} For photoinduced superconductivity,\cite{fausti_s11} electron-phonon interactions are considered to play an essential role, but their effects are still controversial.\cite{babadi_prb17,murakami_prb17} If the coherence among pairs is reduced, conductivity is suppressed.\cite{fukaya_ncomms15} Thus, in many cases, electron motion needs to be driven coherently by an oscillating electric field.\cite{kawakami_prl10,matsubara_prb14} 

Among the photoinduced phenomena, those that are realized after an intense light field is applied to solid-state materials have attracted much attention. Resultant nonequilibrium states are not simply far from equilibrium ones but may show some distinctive dynamic behaviors. High-harmonic generation, which is now being extensively studied experimentally\cite{ghimire_nphys11,schubert_np14,luu_n15,yoshikawa_s17} and theoretically,\cite{silva_np18,murakami_prl18,nag_arx,ikeda_arx} is such a phenomenon, although it appears during photoexcitation and is regarded as a time-periodic steady-state response. In some cases, the synchronized motion of electrons has been suggested to be caused by interactions during\cite{nag_arx} and after\cite{yonemitsu_jpsj18a,kawakami_np18,nag_arx} an intense light field is applied. This fact is reminiscent of the situation in a discrete time crystal,\cite{else_prl16,yao_prl17} which is realized in many-body-localized periodically driven systems:\cite{zhang_n17,choi_n17} interactions are essential for collective synchronization in strongly disordered systems.\cite{yao_prl17} 

Quite recently, a nonlinear charge oscillation and a resultant stimulated emission have been observed in the organic superconductor $\kappa$-(bis[ethylenedithio]tetrathiafulvalene)$_2$Cu[N(CN)$_2$]Br [$\kappa$-(BEDT-TTF)$_2$Cu[N(CN)$_2$]Br],\cite{kawakami_np18} whose lattice has a dimerized structure. This and related materials are known for their phase diagram with dimer-Mott-insulator-metal and superconducting transitions and an unconventional critical behavior.\cite{kagawa_n05} Photoinduced insulator-metal transitions are also  known.\cite{kawakami_prl09,yonemitsu_jpsj11b,gomi_jpsj14} Because of the dimerized structure, its dielectric permittivity shows an anomaly\cite{abdel_prb10} owing to polar charge distributions.\cite{naka_jpsj10,gomi_prb10,hotta_prb10,dayal_prb11,itoh_prl13} Charge fluctuations associated with such distributions may be relevant to the superconducting phase transition.\cite{sekine_prb13,watanabe_jpsj17} 

For the nonlinear charge oscillation to appear, a dimerized structure has been shown to be important by theoretical calculations on the basis of the exact diagonalization method.\cite{yonemitsu_jpsj18a} However, the relationship between this oscillation and superconductivity has not yet been theoretically discussed. Experimentally, it has been observed that the resultant increase in reflectivity is enhanced by superconducting fluctuations and weakened as the temperature decreases below the superconducting transition temperature.\cite{kawakami_np18} Thus, it is important to clarify their relationship theoretically, which would be useful in characterizing the superconducting phase in organic conductors. 

Here, we theoretically study what kind of charge oscillations appear after an intense electric-field pulse is applied to superconducting states on a dimer lattice of a simpler form. To treat the time profiles of charge densities and pairing order parameters, we employ the Hartree-Fock-Gor'kov approximation. In general, without dimerization, time-dependent pairing order parameters have been extensively studied experimentally\cite{matsunaga_s14} and theoretically.\cite{littlewood_prl81,barankov_prl04,yuzbashyan_prl06,tsuji_prb15,sentef_prl17} Dynamic relationships with charge density waves have also been studied.\cite{littlewood_prl81,sentef_prl17} On the other hand, without superconductivity, the synchronized motion of electrons has also been theoretically realized on a honeycomb lattice,\cite{nag_arx} which also has two sites per unit cell. In this paper, we show that novel charge-oscillation modes appear through a dimerization-induced coupling between superconductivity and charge-density modulation after strong photoexcitation. 

\section{Model with On-Site and Intradimer Interactions \label{sec:model}}
\begin{figure}
\includegraphics[height=13.6cm]{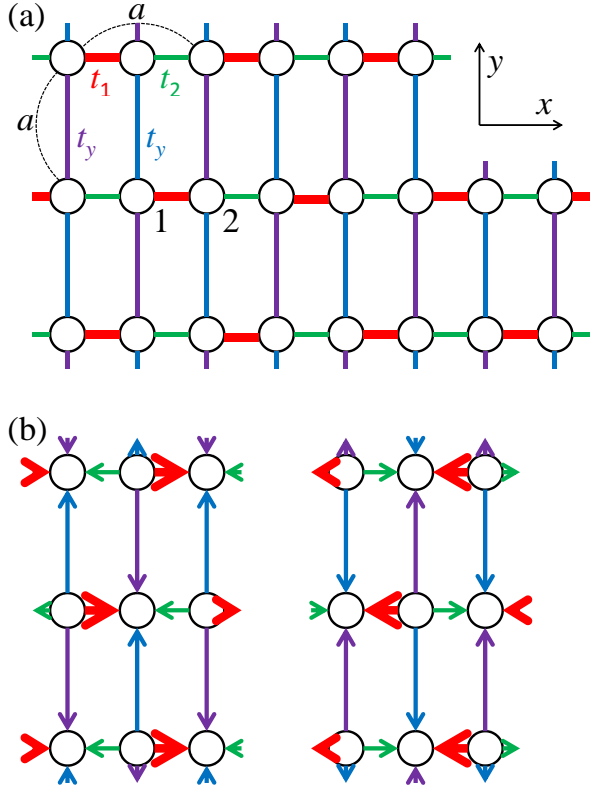}
\caption{(Color online) 
(a) Two-dimensional lattice consisting of dimers. The magnitude of the intradimer transfer integral $t_1$ is larger than those of the interdimer ones $t_2$ and $t_y$. The lattice constant is $a$ along the $x$- and $y$-axes. (b) Current distributions in breathing modes. The left and right distributions alternate during the corresponding charge oscillations. 
\label{fig:dimer_latt}}
\end{figure}

In a previous study,\cite{yonemitsu_jpsj18a} the nonlinear charge oscillation was shown to appear in a one-dimensional spinless-fermion ``$t_1$-$t_2$-$V$'' model at half filling and a two-dimensional extended Hubbard model for $\kappa$-(BEDT-TTF)$_2$X at three-quarter filling. It is briefly mentioned that the appearance is not limited to these fillings. To obtain a hint on how general these models are, which produce nonlinear charge oscillations, and to study the effect of superconductivity on them, we consider a toy model whose lattice structure is much simpler than that of $\kappa$-(BEDT-TTF)$_2$X, 
\begin{equation}
H=H_{\mbox{kin}}+H_{\mbox{int}}
\;, \label{eq:ham}
\end{equation}
where the kinetic term $ H_{\mbox{kin}} $ is a tight-binding model on the dimer lattice shown in Fig.~\ref{fig:dimer_latt}(a) with Peierls phase factors [$ \frac{ea}{\hbar c} A_{x(y)} $ with $ A_{x(y)} $ being the $ x(y) $ component of the vector potential and $ a $ being the lattice constant is simply written as $ A_{x(y)} $ below], 
\begin{eqnarray}
H_{\mbox{kin}} & & = \sum_{ \mbox{\boldmath $i$},\sigma } \left(
t_1 e^{iA_x/2} c^\dagger_{\mbox{\boldmath $i$},1,\sigma} c_{\mbox{\boldmath $i$},2,\sigma} + 
t_1 e^{-iA_x/2} c^\dagger_{\mbox{\boldmath $i$},2,\sigma} c_{\mbox{\boldmath $i$},1,\sigma} \right. \nonumber \\
& & + 
t_2 e^{iA_x/2} c^\dagger_{\mbox{\boldmath $i$},2,\sigma} c_{\mbox{\boldmath $i$}+\mbox{\boldmath $x$},1,\sigma} + 
t_2 e^{-iA_x/2} c^\dagger_{\mbox{\boldmath $i$}+\mbox{\boldmath $x$},1,\sigma} c_{\mbox{\boldmath $i$},2,\sigma} \nonumber \\
& & + 
t_y e^{iA_y} c^\dagger_{\mbox{\boldmath $i$},1,\sigma} c_{\mbox{\boldmath $i$-\mbox{\boldmath $x$}/2+\mbox{\boldmath $y$}},2,\sigma} + 
t_y e^{-iA_y} c^\dagger_{\mbox{\boldmath $i$-\mbox{\boldmath $x$}/2+\mbox{\boldmath $y$}},2,\sigma} c_{\mbox{\boldmath $i$},1,\sigma} \nonumber \\
& & + 
t_y e^{iA_y} c^\dagger_{\mbox{\boldmath $i$},2,\sigma} c_{\mbox{\boldmath $i$+\mbox{\boldmath $x$}/2+\mbox{\boldmath $y$}},1,\sigma} + 
t_y e^{-iA_y} c^\dagger_{\mbox{\boldmath $i$+\mbox{\boldmath $x$}/2+\mbox{\boldmath $y$}},1,\sigma} c_{\mbox{\boldmath $i$},2,\sigma} \nonumber \\
& &  \left. 
-\mu c^\dagger_{\mbox{\boldmath $i$},1,\sigma} c_{\mbox{\boldmath $i$},1,\sigma} 
-\mu c^\dagger_{\mbox{\boldmath $i$},2,\sigma} c_{\mbox{\boldmath $i$},2,\sigma} 
\right) 
\;. \label{eq:kin}
\end{eqnarray}
Here, $ c^\dagger_{\mbox{\boldmath $i$},\alpha,\sigma} $ creates an electron with orbital $ \alpha $ and spin $ \sigma $ at dimer site $ \mbox{\boldmath $i$} $. The intradimer transfer integral $t_1$ and the interdimer ones $t_2$ and $t_y$ are configured as shown in Fig.~\ref{fig:dimer_latt}(a). The parameter $ \mu $ is the chemical potential. Via the Fourier transform $ c^\dagger_{\mbox{\boldmath $j$},\alpha,\sigma} = \frac{1}{\sqrt{N}} \sum_{\mbox{\boldmath $k$}} e^{-i\mbox{\boldmath $k$} \cdot \mbox{\boldmath $j$}} c^\dagger_{\mbox{\boldmath $k$},\alpha,\sigma} $ with $ N $ being the number of dimers, $ H_{\mbox{kin}} $ is rewritten in momentum space as 
\begin{eqnarray}
H_{\mbox{kin}} & & = \sum_{ \mbox{\boldmath $k$} } \left[
e_{12}(\mbox{\boldmath $k$},\mbox{\boldmath $A$}) 
c^\dagger_{\mbox{\boldmath $k$},1,\uparrow} c_{\mbox{\boldmath $k$},2,\uparrow} + 
e_{12}^\ast(\mbox{\boldmath $k$},\mbox{\boldmath $A$}) 
c^\dagger_{\mbox{\boldmath $k$},2,\uparrow} c_{\mbox{\boldmath $k$},1,\uparrow} \right. \nonumber \\
& & 
-\mu \left( c^\dagger_{\mbox{\boldmath $k$},1,\uparrow} c_{\mbox{\boldmath $k$},1,\uparrow} + c^\dagger_{\mbox{\boldmath $k$},2,\uparrow} c_{\mbox{\boldmath $k$},2,\uparrow} \right) \nonumber \\
& & 
-e_{12}^\ast(-\mbox{\boldmath $k$},\mbox{\boldmath $A$}) 
c_{-\mbox{\boldmath $k$},1,\downarrow} c^\dagger_{-\mbox{\boldmath $k$},2,\downarrow} 
-e_{12}(-\mbox{\boldmath $k$},\mbox{\boldmath $A$}) 
c_{-\mbox{\boldmath $k$},2,\downarrow} c^\dagger_{-\mbox{\boldmath $k$},1,\downarrow} \nonumber \\
& & \left. 
+\mu \left( c_{-\mbox{\boldmath $k$},1,\downarrow} c^\dagger_{-\mbox{\boldmath $k$},1,\downarrow} + c_{-\mbox{\boldmath $k$},2,\downarrow} c^\dagger_{-\mbox{\boldmath $k$},2,\downarrow} -2 \right) 
\right]
\;, \label{eq:kin_mom}
\end{eqnarray}
where $ e_{12}(\mbox{\boldmath $k$},\mbox{\boldmath $A$}) $ is defined as 
\begin{equation}
e_{12}(\mbox{\boldmath $k$},\mbox{\boldmath $A$}) = 
t_1 e^{iA_x/2} + t_2 e^{-ik_x-iA_x/2} + 2 t_y e^{-ik_x/2} \cos(k_y+A_y) 
\;. \label{eq:e_12}	
\end{equation}

For the interaction term $ H_{\mbox{int}} $, we consider on-site and intradimer interactions, 
\begin{eqnarray}
H_{\mbox{int}} & & = \sum_{\mbox{\boldmath $i$}} \left[ 
U_1 n_{\mbox{\boldmath $i$},1,\uparrow} n_{\mbox{\boldmath $i$},1,\downarrow} + 
U_2 n_{\mbox{\boldmath $i$},2,\uparrow} n_{\mbox{\boldmath $i$},2,\downarrow} + 
\right. \nonumber \\
& & 
+2V_{\mbox{sp}} \mbox{\boldmath $S$}_{\mbox{\boldmath $i$},1} \cdot 
\mbox{\boldmath $S$}_{\mbox{\boldmath $i$},2} 
-\frac{V_{\mbox{ch}}}{2} n_{\mbox{\boldmath $i$},1} 
n_{\mbox{\boldmath $i$},2} 
\nonumber \\
& & \left.
+V_{\mbox{ph}} \left(
\Delta^\dagger_{\mbox{\boldmath $i$},11} 
\Delta_{\mbox{\boldmath $i$},22} + 
\Delta^\dagger_{\mbox{\boldmath $i$},22} 
\Delta_{\mbox{\boldmath $i$},11} 
\right) \right]
\;, \label{eq:int}
\end{eqnarray}
where 
$ n_{\mbox{\boldmath $i$},\alpha,\sigma} = 
c^\dagger_{\mbox{\boldmath $i$},\alpha,\sigma} 
c_{\mbox{\boldmath $i$},\alpha,\sigma} $, 
$ \mbox{\boldmath $S$}_{\mbox{\boldmath $i$},\alpha} = \frac12 \sum_{\tau,\tau'} 
c^\dagger_{\mbox{\boldmath $i$},\alpha,\tau} 
\mbox{\boldmath $\sigma$}_{\tau,\tau'}
c_{\mbox{\boldmath $i$},\alpha,\tau'}
$ with $ \mbox{\boldmath $\sigma$} $ being the Pauli matrices, 
$ n_{\mbox{\boldmath $i$},\alpha} = \sum_\sigma 
n_{\mbox{\boldmath $i$},\alpha,\sigma} $, and 
$ \Delta_{\mbox{\boldmath $i$},\alpha\beta} = 
c_{\mbox{\boldmath $i$},\alpha,\downarrow} 
c_{\mbox{\boldmath $i$},\beta,\uparrow} $. 
In this paper, we use the on-site attraction $ U_1 = U_2 = U < 0 $ to induce $ s $-wave pairing. The interorbital spin-spin (charge-charge) interaction $ V_{\mbox{sp}} $ ($ V_{\mbox{ch}} $) and the pair-hopping interaction $ V_{\mbox{ph}} $ are added to study their effects on charge oscillations in $ s $-wave superconductivity. The $ V_{\mbox{sp}} $ and $ V_{\mbox{ch}} $ terms were studied in Ref.~\citen{bishop_prb16} and found to induce interorbital pairing for $ V_{\mbox{sp}} = V_{\mbox{ch}} > 0 $ with no other interactions. The network of transfer integrals was chosen in such a way that the pairing was $ d $-wave in Ref.~\citen{bishop_prb16}, but this is not the case in the present paper. 

We employ the Hartree-Fock-Gor'kov approximation and replace $ H_{\mbox{int}} $ by $ H_{\mbox{int}}^{\mbox{MF}} $, assuming 
\begin{equation}
\langle c^\dagger_{\mbox{\boldmath $i$},\alpha,\sigma} 
c_{\mbox{\boldmath $i$},\alpha,\sigma} \rangle = n_{\alpha,\sigma} 
\;, \label{eq:def_n}
\end{equation}
\begin{equation}
\langle c^\dagger_{\mbox{\boldmath $i$},1,\sigma} 
c_{\mbox{\boldmath $i$},2,\sigma} \rangle = f_{\sigma} 
\;, \label{eq:def_f}
\end{equation}
and 
\begin{equation}
\langle c_{\mbox{\boldmath $i$},\alpha,\downarrow} 
c_{\mbox{\boldmath $i$},\beta,\uparrow} \rangle = \Delta_{\alpha\beta} 
\;. \label{eq:def_d}
\end{equation}
Using 
$ V_{\pm} \equiv \frac{1}{2} \left( V_{\mbox{sp}} \pm V_{\mbox{ch}} \right)  $, 
$ \bar{\alpha}=2 $ for $ \alpha=1 $, and $ \bar{\alpha}=1 $ for $ \alpha=2 $, $ H_{\mbox{int}}^{\mbox{MF}} $ is written as 
\begin{equation}
H_{\mbox{int}}^{\mbox{MF}} = H_{\mbox{d}} 
+ H_{\mbox{o}} 
+ H_{\mbox{p}} 
+ H_{\mbox{c}} 
\;, \label{eq:MF}
\end{equation}
where 
\begin{eqnarray}
H_{\mbox{d}} & & = \sum_{\mbox{\boldmath $k$},\alpha} \left[ 
\left( U_{\alpha} n_{\alpha,\downarrow} 
-V_{+} n_{\bar{\alpha},\downarrow} 
+V_{-} n_{\bar{\alpha},\uparrow}  \right) 
c^\dagger_{\mbox{\boldmath $k$},\alpha,\uparrow} 
c_{\mbox{\boldmath $k$},\alpha,\uparrow} \right. \nonumber \\
& & \left. - 
\left( U_{\alpha} n_{\alpha,\uparrow} 
-V_{+} n_{\bar{\alpha},\uparrow} 
+V_{-} n_{\bar{\alpha},\downarrow} \right) 
c_{-\mbox{\boldmath $k$},\alpha,\downarrow} 
c^\dagger_{-\mbox{\boldmath $k$},\alpha,\downarrow} \right]
\;, \label{eq:int_d}
\end{eqnarray}
\begin{eqnarray}
H_{\mbox{o}} & & = \sum_{\mbox{\boldmath $k$}} \left[ 
-\left( V_{-} f^\ast_{\uparrow} 
+ V_{\mbox{sp}} f^\ast_{\downarrow} 
- V_{\mbox{ph}} f_{\downarrow} \right) 
c^\dagger_{\mbox{\boldmath $k$},1,\uparrow} 
c_{\mbox{\boldmath $k$},2,\uparrow} \right. \nonumber \\
& & \left. 
+\left( V_{-} f_{\downarrow} 
+ V_{\mbox{sp}} f_{\uparrow} 
- V_{\mbox{ph}} f^\ast_{\uparrow} \right) 
c_{-\mbox{\boldmath $k$},1,\downarrow} 
c^\dagger_{-\mbox{\boldmath $k$},2,\downarrow} \right] \nonumber \\
& & +\mbox{H.c.} 
\;, \label{eq:int_o}
\end{eqnarray}
\begin{eqnarray}
H_{\mbox{p}} & & = \sum_{\mbox{\boldmath $k$},\alpha} \left[ 
\left( U_{\alpha} \Delta_{\alpha\alpha} 
+ V_{\mbox{ph}} \Delta_{\bar{\alpha}\bar{\alpha}} \right) 
c^\dagger_{\mbox{\boldmath $k$},\alpha,\uparrow} 
c^\dagger_{-\mbox{\boldmath $k$},\alpha,\downarrow} \right. \nonumber \\
& & \left. 
-\left( V_{\mbox{sp}} \Delta_{\alpha\bar{\alpha}} 
+ V_{+} \Delta_{\bar{\alpha}\alpha} \right) 
c^\dagger_{\mbox{\boldmath $k$},\alpha,\uparrow} 
c^\dagger_{-\mbox{\boldmath $k$},\bar{\alpha},\downarrow} \right] \nonumber \\
& & +\mbox{H.c.}
\;, \label{eq:int_p}
\end{eqnarray}
and 
\begin{eqnarray}
H_{\mbox{c}} & & = N \left\{ 
\sum_{\alpha} \left[ \left( U_{\alpha}-V_{+} \right) n_{\alpha,\uparrow} 
+V_{-} n_{\alpha,\downarrow} \right] \right. \nonumber \\
& & -\sum_{\alpha} U_{\alpha} 
\left( n_{\alpha,\uparrow} n_{\alpha,\downarrow} + 
\mid \Delta_{\alpha\alpha} \mid^2 \right) \nonumber \\
& & + V_{+} \sum_{\alpha} \left( 
n_{\alpha,\uparrow} n_{\bar{\alpha},\downarrow} + 
\mid \Delta_{\alpha\bar{\alpha}} \mid^2 \right) \nonumber \\
& & + V_{-} \sum_{\sigma} \left( 
-n_{1,\sigma} n_{2,\sigma} + \mid f_{\sigma} \mid^2 \right) \nonumber \\
& & + V_{\mbox{sp}} \left( 
\sum_{\sigma} f^\ast_{\sigma} f_{\bar{\sigma}} + \sum_{\alpha} 
\Delta^\ast_{\alpha\bar{\alpha}} \Delta_{\bar{\alpha}\alpha} \right) \nonumber \\
& & \left. - V_{\mbox{ph}} \left( 
f_{\uparrow} f_{\downarrow} + f^\ast_{\uparrow} f^\ast_{\downarrow} +\sum_{\alpha} 
\Delta^\ast_{\alpha\alpha} \Delta_{\bar{\alpha}\bar{\alpha}} \right) \right\} 
\;. \label{eq:int_c}
\end{eqnarray}
The order parameters are calculated using the mean-field Hamiltonian $ H^{\mbox{MF}} = H_{\mbox{kin}} + H_{\mbox{int}}^{\mbox{MF}} $ with the formulae 
\begin{equation}
n_{\alpha,\sigma} = \frac{1}{N} \sum_{\mbox{\boldmath $k$}} 
\langle c^\dagger_{\mbox{\boldmath $k$},\alpha,\sigma} 
c_{\mbox{\boldmath $k$},\alpha,\sigma} \rangle 
\;, \label{eq:op_n}
\end{equation}
\begin{equation}
f_{\sigma} = \frac{1}{N} \sum_{\mbox{\boldmath $k$}} 
\langle c^\dagger_{\mbox{\boldmath $k$},1,\sigma} 
c_{\mbox{\boldmath $k$},2,\sigma} \rangle 
\;, \label{eq:op_f}
\end{equation}
and 
\begin{equation}
\Delta_{\alpha\beta} = \frac{1}{N} \sum_{\mbox{\boldmath $k$}} 
\langle c_{-\mbox{\boldmath $k$},\alpha,\downarrow} 
c_{\mbox{\boldmath $k$},\beta,\uparrow} \rangle 
\;, \label{eq:op_d}
\end{equation}
which are numerically iterated until convergence in obtaining the ground state. For the model parameters, we use $ t_1 = -0.3 $, $ t_2 = t_y = -0.1 $, and $ V_{\mbox{sp}} = V_{\mbox{ch}} = V_{\mbox{ph}} = 0 $ unless stated otherwise. The chemical potential $ \mu $ is set so that the system is at three-quarter filling (i.e., three electrons per dimer) throughout this paper. Even at half filling, which is insulating without interactions, nonlinear charge oscillations appear. To study the effect of superconductivity on them, however, the filling is set to be away from half filling. For the system size, we use $N$=100$\times$100 and periodic boundary conditions. This is sufficiently large for the spectra and time profiles shown in this paper: the finite-size effects are smaller than the symbols. 

The initial state is the mean-field ground state. For $ U<0 $, it is always a superconducting state. Photoexcitation is introduced through the Peierls phase, where different time-dependent vector potentials are used depending on the purpose. For absorption spectra, we use slowly decaying oscillating electric fields with a small amplitude,\cite{tanaka_jpsj10} 
\begin{equation}
\mbox{\boldmath $A$} (t) = 
\frac{\mbox{\boldmath $F$}\theta(t)}{\omega^2+\gamma^2} 
\left\{ e^{-\gamma t} \left[ 
\omega \cos(\omega t) + \gamma \sin(\omega t) \right] -\omega \right\}
\;, \label{eq:for_abs}
\end{equation}
where $ \omega $ is the frequency and $ \gamma $ is the decay constant. 
For photoinduced charge oscillations and associated Fourier spectra, we use symmetric one-cycle electric-field pulses,\cite{yonemitsu_jpsj15,yanagiya_jpsj15,yonemitsu_jpsj17a,yonemitsu_jpsj18a} 
\begin{equation}
\mbox{\boldmath $A$} (t) = \frac{\mbox{\boldmath $F$}}{\omega_c} 
\theta (t) \theta \left( \frac{2\pi}{\omega_c}-t \right) 
\left[ \cos (\omega_c t)-1 \right] 
\;, \label{eq:pulse}
\end{equation}
where the central frequency $ \omega_c $ is chosen to be $ \omega_c = 0.7 $ throughout the paper because the qualitative results are independent of its value as in the previous study.\cite{yonemitsu_jpsj18a} In both cases, the maximum electric field $ \mbox{\boldmath $F$} $ is polarized along $ (1, 1) $, $ \mbox{\boldmath $F$} = (F, F) $, unless stated otherwise. The time-dependent Schr\"odinger equation is numerically solved by directly extending the method that uses the Hartree-Fock approximation.\cite{terai_ptp93,kuwabara_jpsj95,miyashita_jpsj03,tanaka_prb18} 

In contrast to the previous study,\cite{yonemitsu_jpsj18a} every dimer is always equivalent for any polarization of $ \mbox{\boldmath $F$} $ as assumed in Eqs.~(\ref{eq:def_n})--(\ref{eq:def_d}). In the ground state, the system satisfies $ \sum_\sigma n_{1,\sigma} = \sum_\sigma n_{2,\sigma} $. Photoexcitation with $ F_x \neq 0 $ leads to $ \sum_\sigma n_{1,\sigma} \neq \sum_\sigma n_{2,\sigma} $. Thus, the time profile of $ \sum_\sigma n_{1,\sigma} $ (or $ \sum_\sigma n_{2,\sigma} $) has information that is shared with the optical conductivity or absorption spectrum. Then, as in the previous study,\cite{yonemitsu_jpsj18a} we calculate the absolute values of the Fourier transforms of these time profiles and refer to them as Fourier spectra. The time span used for the Fourier spectra is mainly $ T < t < 50T $ with $ T = 2\pi/\omega_c $, but we use $ T < t < 500T $ when oscillation frequencies with a higher resolution are required. 

To facilitate the characterization of charge oscillations, we calculate current order parameters, which are defined as 
\begin{equation}
j_{x1}= \sum_{\sigma} \left( 
-i e^{iA_x/2} c^\dagger_{\mbox{\boldmath $i$},1,\sigma} c_{\mbox{\boldmath $i$},2,\sigma} 
+i e^{-iA_x/2} c^\dagger_{\mbox{\boldmath $i$},2,\sigma} c_{\mbox{\boldmath $i$},1,\sigma} \right) 
\;, \label{eq:jx1}
\end{equation}
\begin{equation}
j_{x2}= \sum_{\sigma} \left( 
-i e^{iA_x/2} c^\dagger_{\mbox{\boldmath $i$},2,\sigma} c_{\mbox{\boldmath $i$}+\mbox{\boldmath $x$},1,\sigma} 
+i e^{-iA_x/2} c^\dagger_{\mbox{\boldmath $i$}+\mbox{\boldmath $x$},1,\sigma} c_{\mbox{\boldmath $i$},2,\sigma} \right) 
\;, \label{eq:jx2}
\end{equation}
\begin{equation}
j_{y1}= \sum_{\sigma} \left( 
-i e^{iA_y} c^\dagger_{\mbox{\boldmath $i$},1,\sigma} c_{\mbox{\boldmath $i$-\mbox{\boldmath $x$}/2+\mbox{\boldmath $y$}},2,\sigma} 
+i e^{-iA_y} c^\dagger_{\mbox{\boldmath $i$-\mbox{\boldmath $x$}/2+\mbox{\boldmath $y$}},2,\sigma} c_{\mbox{\boldmath $i$},1,\sigma} \right) 
\;, \label{eq:jy1}
\end{equation}
and 
\begin{equation}
j_{y2}= \sum_{\sigma} \left( 
-i e^{iA_y} c^\dagger_{\mbox{\boldmath $i$},2,\sigma} c_{\mbox{\boldmath $i$+\mbox{\boldmath $x$}/2+\mbox{\boldmath $y$}},1,\sigma} 
+i e^{-iA_y} c^\dagger_{\mbox{\boldmath $i$+\mbox{\boldmath $x$}/2+\mbox{\boldmath $y$}},1,\sigma} c_{\mbox{\boldmath $i$},2,\sigma} \right) 
\;, \label{eq:jy2}
\end{equation}
which are independent of $ \mbox{\boldmath $i$} $. They are related to the current density $ \mbox{\boldmath $j$} = -(1/N)\partial H/\partial \mbox{\boldmath $A$} $ through 
\begin{equation}
\langle j_x \rangle = (t_1/2) j_{x1} + (t_2/2) j_{x2} 
\;, \;\;\;\;\; 
\langle j_y \rangle = t_y \left( j_{y1} + j_{y2} \right) 
\;. \label{eq:cur}
\end{equation}

\section{Photoinduced Charge Oscillations}

\subsection{Absorption spectra}
\begin{figure}
\includegraphics[height=13.6cm]{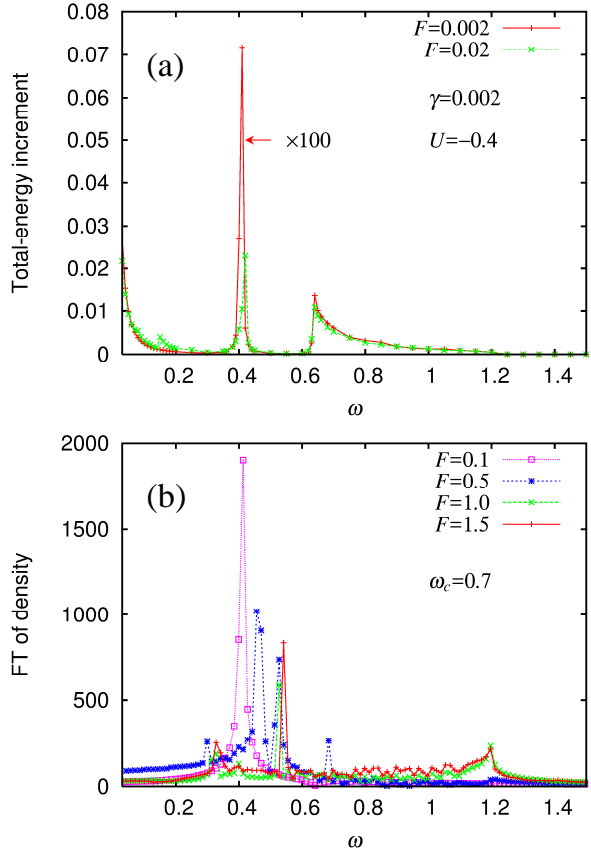}
\caption{(Color online) 
(a) Increase in total energy after application of electric field with exponential decay constant $ \gamma = 0.002 $ and field amplitude $ F=0.002 $ (multiplied by a factor of 100), compared with that with $ F = 0.02 $. (b) Absolute values of Fourier transforms of time profiles ($ T < t < 50T $) of charge density after application of one-cycle electric-field pulses with central frequency $ \omega_c = 0.7 $ and different field amplitudes. The model parameters are $ t_1 = -0.3 $, $ t_2 = t_y = -0.1 $, and $ U = -0.4 $. 
\label{fig:las_d_n3_311_um4}}
\end{figure}

The mean-field Hamiltonian $ H^{\mbox{MF}} $ is described by a $ 4 \times 4 $ matrix at each $ \mbox{\boldmath $k$} $, so that two bands are located above the chemical potential and two bands are below it. 
For absorption spectra in the ground state with $ U=-0.4 $, we calculate the increase in the total energy after the application of an electric field with $ \gamma = 0.002 $ and $ F=0.002 $ for different $ \omega $ and show them in Fig.~\ref{fig:las_d_n3_311_um4}(a). 
The peak around $ \omega \simeq 0.4 $ is due to transitions from the second (in the order of lowest energy to highest energy) band to the third band. The absorption band in the range of $ 0.6 < \omega < 1.2 $ is due to transitions from the first (second) to third (fourth) bands. This absorption band exists in the normal phase with $ U = 0 $, which originates from dimerization-induced interband transitions from bonding to antibonding states. 

Because the Higgs mode is not excited in the first order with respect to $ \mbox{\boldmath $A$} $,\cite{tsuji_prb15} it does not appear in a linear absorption spectrum. Then, we make the field amplitude ten times larger ($ F=0.02 $) and show the resultant spectrum in Fig.~\ref{fig:las_d_n3_311_um4}(a). Because the energy supplied by the field basically becomes 100 times larger, the spectrum for $ F=0.002 $ is multiplied by 100 for comparison. Since the difference between the lowest unoccupied and highest occupied energy levels is 0.153, the so-called Higgs mode is responsible for the peak at $ \omega = 0.153 $ for the larger $ F $. In fact, for the initial state without imaginary parts of $ \Delta_{\alpha\alpha} $, the photoexcitation with this frequency oscillates not only the real parts of $ \Delta_{11} $ and $ \Delta_{22} $ (in phase) but also their imaginary parts (antiphase) and the charge-density difference because the latter is coupled to the pairing order parameters for nonzero dimerization. Because the imaginary parts of $ \Delta_{\alpha\alpha} $ are smaller than the real parts, their magnitudes mainly oscillate with this frequency. In this sense, this mode is regarded as the Higgs mode. However, we will no longer discuss the Higgs mode since the absorption due to the Higgs mode is generally very weak and it barely affects charge oscillations. 

Note that, without dimerization ($ \mid t_1 \mid = \mid t_2 \mid $), $ \mid \Delta_{\alpha\alpha} \mid $ oscillates (although its amplitude is generally small), but $ n_{\alpha,\sigma} $ remains constant after photoexcitation. Nonzero dimerization ($ \mid t_1 \mid > \mid t_2 \mid $) makes $ n_{\alpha,\sigma} $ oscillate after photoexcitation and increases the oscillation amplitude of $ \mid \Delta_{\alpha\alpha} \mid $ through the interband transitions from bonding to antibonding quasiparticle states, which obscure the Higgs mode. 

\subsection{Field-amplitude dependence}
Hereafter, we show Fourier spectra of charge densities ($ \sum_\sigma n_{\alpha,\sigma} $) after photoexcitation. Those for $ U=-0.4 $ are shown with different $ F $ in Fig.~\ref{fig:las_d_n3_311_um4}(b). For small $ F $ ($ F=0.1 $), it has spectral weights at energies where the linear absorption spectrum [corresponding to the smaller-$ F $ case in Fig.~\ref{fig:las_d_n3_311_um4}(a) except for the tail continuing from zero energy due to $ \gamma > 0 $] has spectral weights. The lower-energy peak at $ \omega \simeq 0.4 $ is conspicuous, and small but nonzero weights exist in the range of $ 0.6 < \omega < 1.2 $. Because $ \mid \Delta_{\alpha\alpha} \mid $ is coupled to $ \sum_\sigma n_{1,\sigma} - \sum_\sigma n_{2,\sigma} $ for nonzero dimerization, $ \mid \Delta_{\alpha\alpha} \mid $ also oscillates with $ \omega \simeq 0.4 $. The conspicuous peak becomes slightly blueshifted and lowered with increasing $ F $ (from $ F=0.1 $ to 0.5). It is suppressed for large $ F $. 

For large $ F $ ($ F=1.0 $, 1.5), new charge-oscillation modes appear and their frequencies are not basically shifted (compare those for $ F=1.0 $ with those for $ F=1.5 $). In the present case with $ U=-0.4 $, their frequencies are $ \omega=0.33 $, 0.54, and 1.2. That of $ \omega=1.2 $ appears even in the noninteracting case. It is referred to as a high-frequency charge-oscillation mode. That of $ \omega=0.54 $ also appears in the Fourier spectrum of $ \mid \Delta_{\alpha\alpha} \mid $ (not shown); thus, $ \mid \Delta_{\alpha\alpha} \mid $ is coupled to $ \sum_\sigma n_{1,\sigma} - \sum_\sigma n_{2,\sigma} $ in this mode. It is referred to as a middle-frequency charge-oscillation mode. These modes are analyzed later in detail. The charge-oscillation mode of $ \omega=0.33 $ also appears in the Fourier spectrum of $ \mid \Delta_{\alpha\bar{\alpha}} \mid $ (not shown); thus, $ \mid \Delta_{\alpha\bar{\alpha}} \mid $ is coupled to $ \sum_\sigma n_{1,\sigma} - \sum_\sigma n_{2,\sigma} $ in this mode. Its weight in the Fourier spectrum of $ \sum_\sigma n_{\alpha,\sigma} $ is relatively small as long as the on-site attraction $ \mid U \mid $ is a dominant interaction. The main results for large $ F $ here and below are independent of the polarization of $ \mbox{\boldmath $F$} $ as long as $ F_x \neq 0 $. This is in contrast to the two-dimensional case in the previous study,\cite{yonemitsu_jpsj18a} where every dimer becomes equivalent only for specifically polarized fields. 

\subsection{Interaction-strength dependence}
\begin{figure}
\includegraphics[height=20.4cm]{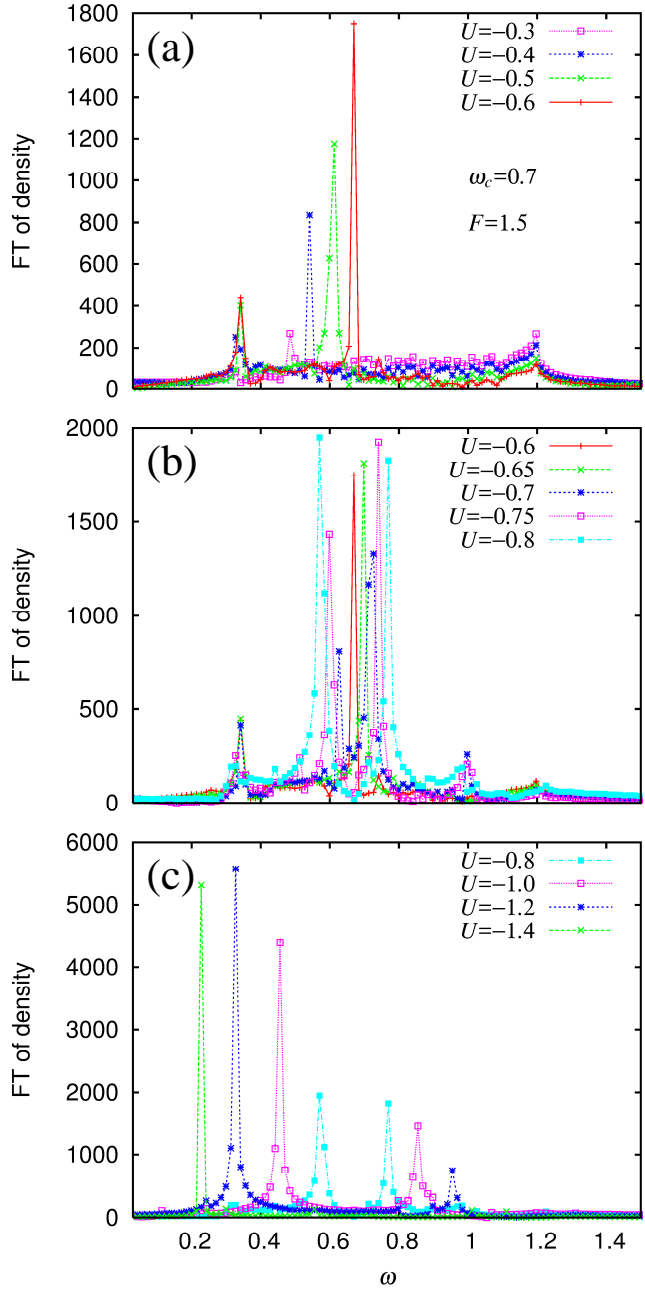}
\caption{(Color online) 
Absolute values of Fourier transforms of time profiles ($ T < t < 50T $) of charge densities after application of one-cycle electric-field pulse with $ \omega_c=0.7 $ and $ F=1.5 $ for (a) weak, (b) intermediate, and (c) strong on-site attractions. The other model parameters are $ t_1 = -0.3 $ and $ t_2 = t_y = -0.1 $. 
\label{fig:d_n3_311_umx_w7_e1p5FT}}
\end{figure}

Hereafter, we focus on charge-oscillation modes that appear for large $ F $ and observe their behaviors with different model parameters. Fourier spectra for weak, intermediate, and strong on-site attractions are shown in Figs.~\ref{fig:d_n3_311_umx_w7_e1p5FT}(a)--\ref{fig:d_n3_311_umx_w7_e1p5FT}(c), respectively. 
For weak interactions, the high-frequency charge-oscillation mode always appears at $ \omega=1.2 $. Its peak height decreases with increasing $ \mid U \mid $. For the middle-frequency charge-oscillation mode, its frequency and peak height increase with $ \mid U \mid $ as shown in Fig.~\ref{fig:d_n3_311_umx_w7_e1p5FT}(a). The charge-oscillation mode coupled to $ \mid \Delta_{\alpha\bar{\alpha}} \mid $ appears at $ \omega=0.33 $ for weak to intermediate on-site attractions. However, it does not dominate the Fourier spectra of $ \sum_\sigma n_{\alpha,\sigma} $. This mode will not be discussed hereafter. 

\begin{figure}
\includegraphics[height=13.6cm]{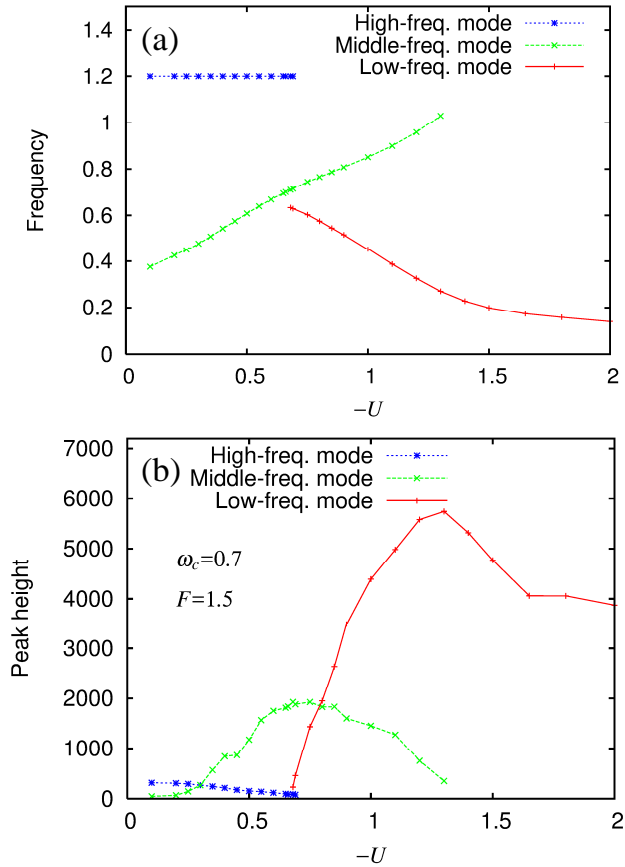}
\caption{(Color online) 
(a) Frequencies and (b) peak heights of three charge-oscillation modes appearing in Fourier spectra as functions of on-site attraction $ \mid U \mid $. The other parameters are $ \omega_c=0.7 $, $ F=1.5 $, $ t_1 = -0.3 $, and $ t_2 = t_y = -0.1 $. For (a), the time span $ T < t < 500T $ is used for the Fourier spectra with higher resolution. 
\label{fig:thr_n3_311_umx}}
\end{figure}

For intermediate on-site attractions, the high-frequency charge-oscillation mode is almost invisible. In addition to the middle-frequency charge-oscillation mode, whose frequency increases with $ \mid U \mid $ and peak height shows a maximum as a function of $ \mid U \mid $, a new mode appears on its low-energy side in Fig.~\ref{fig:d_n3_311_umx_w7_e1p5FT}(b). It is referred to as a low-frequency charge-oscillation mode. For this mode, its frequency decreases and its peak height steeply increases from zero as $ \mid U \mid $ increases; thus, this mode is {\em not} split from the middle-frequency charge-oscillation mode. The interaction strength where the peak height of the middle-frequency charge-oscillation mode becomes maximum is almost identical with the interaction strength where the low-frequency charge-oscillation mode appears ($ U=-0.68 $). The frequencies and peak heights of the three charge-oscillation modes are shown in Figs.~\ref{fig:thr_n3_311_umx}(a) and \ref{fig:thr_n3_311_umx}(b), respectively, as functions of $ \mid U \mid $. 

For strong on-site attractions, only the middle-frequency and low-frequency charge-oscillation modes are visible in the energy range shown in Fig.~\ref{fig:d_n3_311_umx_w7_e1p5FT}(c). Here, their frequencies are separated further as $ \mid U \mid $ increases. The low-frequency charge-oscillation mode is now dominant, and its peak height shows a maximum as a function of $ \mid U \mid $. Note that the present mean-field approximation always produces narrow peaks in Fourier spectra, whose widths are comparable to the frequency slice; thus, the peak heights shown in Fig.~\ref{fig:thr_n3_311_umx}(b) should be understood as a guide. In the following subsections, we show the characteristics of the charge-oscillation modes in Fig.~\ref{fig:thr_n3_311_umx} one by one. 

\subsection{High-frequency charge-oscillation mode}
\begin{figure}
\includegraphics[height=13.6cm]{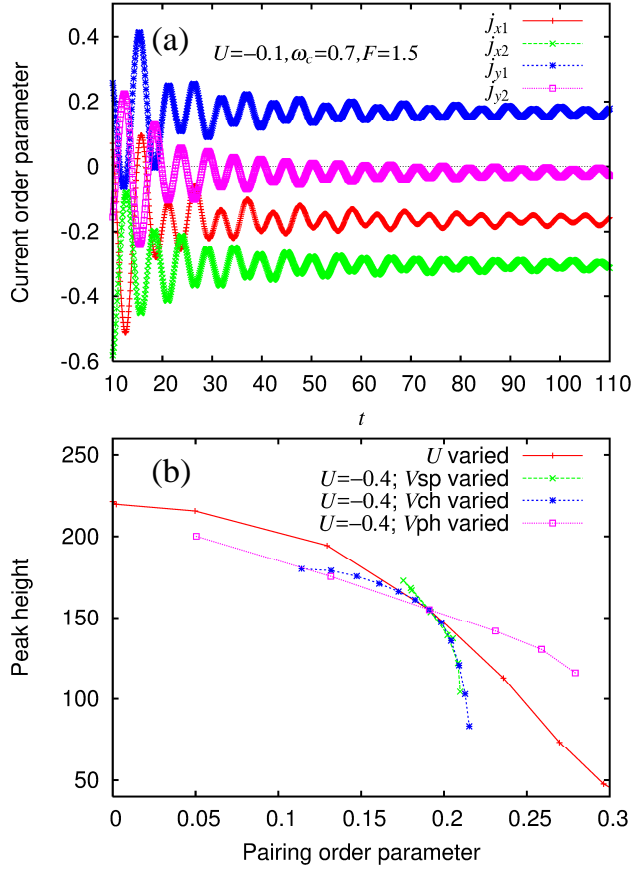}
\caption{(Color online) 
(a) Time profiles of current order parameters, $ j_{x1} $, $ j_{x2} $, $ j_{y1} $, and $ j_{y2} $, for weak on-site attraction $ U = -0.1 $. (b) Peak height of high-frequency charge-oscillation mode as a function of pairing order parameter $ \mid \Delta_{\alpha\alpha} \mid $ in ground state with different on-site attractions. Those for $ U = -0.4 $ with varying $ V_{\mbox{sp}} $, $ V_{\mbox{ch}} $, or $ V_{\mbox{ph}} $ are also shown. The other parameters are $ \omega_c=0.7 $, $ F=1.5 $, $ t_1 = -0.3 $, and $ t_2 = t_y = -0.1 $. 
\label{fig:c_um1_and_d_umxum4xxx}}
\end{figure}

For $ U=-0.1 $, where the high-frequency charge-oscillation mode is dominant in the Fourier spectrum, the time profiles of the current order parameters defined in Sect.~\ref{sec:model} are shown in Fig.~\ref{fig:c_um1_and_d_umxum4xxx}(a). 
Their colors (darkness) hereafter correspond to those of the bonds and the arrows in Fig.~\ref{fig:dimer_latt}. 
Note that their plots are not shifted vertically: their centers of oscillations are nonzero. The system size used here is sufficiently large to ensure that this is not a finite-size effect. The current density induced by the first half of the one-cycle electric-field pulse ($j_{x1}, j_{x2}, j_{y1}, j_{y2} < 0 $) is not completely canceled by its second half even after time averaging. This means that the electrons drift after photoexcitation. The drift of electrons occurs only for weak interactions. As shown later, the centers of oscillations of the current order parameters are zero for intermediate to strong interactions. Here, the oscillating components of $ j_{x1} $ and $ j_{y1} $ are opposite in sign to those of $ j_{x2} $ and $ j_{y2} $. Thus, the current order parameters oscillate back and forth between the left and right panels of Fig.~\ref{fig:dimer_latt}(b). This mode was called an electronic breathing mode in the previous study.\cite{yonemitsu_jpsj18a} These patterns of current distributions maximize the charge-density difference $ \mid \sum_\sigma n_{1,\sigma} - \sum_\sigma n_{2,\sigma} \mid $. As derived in the previous study,\cite{yonemitsu_jpsj18a} the frequency of the electronic breathing mode is given by the relation 
\begin{equation}
\omega_{\mbox{osc}} = 2\left( \mid t_1 \mid + \mid t_2 \mid +2 \mid t_y \mid \right) 
\;, \label{eq:w_osc}
\end{equation}
which is independent of $ U $, and takes a value of 1.2 in the present case. 

In Fig.~\ref{fig:thr_n3_311_umx}(b), the peak height of the electronic breathing mode is shown to decrease with increasing $ \mid U \mid $. This fact implies that the charge-density difference $ \mid \sum_\sigma n_{1,\sigma} - \sum_\sigma n_{2,\sigma} \mid $ competes with the pairing order parameter $ \mid \Delta_{\alpha\alpha} \mid $ during the charge oscillation. We vary $ U $ and plot the peak height as a function of $ \mid \Delta_{\alpha\alpha} \mid $ in Fig.~\ref{fig:c_um1_and_d_umxum4xxx}(b). In addition, we fix $ U=-0.4 $, vary either $ V_{\mbox{sp}} $, $ V_{\mbox{ch}} $, or $ V_{\mbox{ph}} $ from zero, and plot the peak height in the same figure. Note that $ \mid \Delta_{\alpha\alpha} \mid $ is increased in the cases of $ V_{\mbox{sp}} > 0 $ (antiferromagnetic spin-spin interaction), $ V_{\mbox{ch}} > 0 $ (attractive charge-charge interaction), and $ V_{\mbox{ph}} < 0 $ (where $ \Delta_{11} $ cooperates with $ \Delta_{22} $). In all cases, the peak height of the electronic breathing mode decreases as $ \mid \Delta_{\alpha\alpha} \mid $ increases, which makes the competition clear. 

\subsection{Low-frequency charge-oscillation mode}
\begin{figure}
\includegraphics[height=13.6cm]{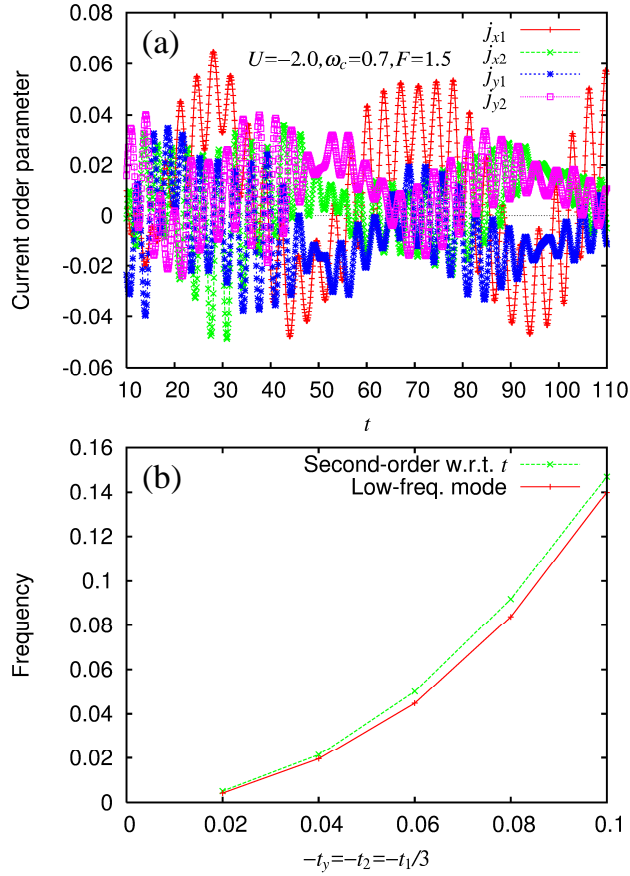}
\caption{(Color online) 
(a) Time profiles of current order parameters, $ j_{x1} $, $ j_{x2} $, $ j_{y1} $, and $ j_{y2} $, for strong on-site attraction $ U = -2.0 $ with $ t_1 = -0.3 $ and $ t_2 = t_y = -0.1 $. (b) Frequency of low-frequency charge-oscillation mode as a function of $ \mid t_y \mid $ with $ t_y=t_2=t_1/3 $, for $ U=-2.0 $. The other parameters are $ \omega_c=0.7 $ and $ F=1.5 $. For (b), the time span $ T < t < 500T $ is used for the Fourier spectra with higher resolution. Also shown is the second-order estimation with respect to transfer integrals [Eq.~(\ref{eq:w_pair_MF_osc})], which is explained in the text. 
\label{fig:c_um2p0_and_11t4500_3xxx}}
\end{figure}

For $ U=-2.0 $, where the low-frequency charge-oscillation mode is dominant in the Fourier spectrum, the time profiles of the current order parameters are shown in Fig.~\ref{fig:c_um2p0_and_11t4500_3xxx}(a). 
They have a slowly varying component and a rapidly varying component. The rapidly varying component is almost invisible in the Fourier spectrum of $ \sum_\sigma n_{\alpha,\sigma} $ even if the frequency range is chosen appropriately. This is because the rapidly varying components of $ j_{x1} $ and $ j_{y2} $ are opposite in sign to those of $ j_{x2} $ and $ j_{y1} $ (i.e., if a charge flows into a site from the $\pm x$ directions, it flows out in the $\pm y$ directions) so as not to affect the charge distribution. On the other hand, the slowly varying component dominates the Fourier spectrum. The slowly varying components of $ j_{x1} $ and $ j_{y1} $ are opposite in sign to those of $ j_{x2} $ and $ j_{y2} $ (i.e., a charge either flows into a site from all directions or flows out in all directions) so as to maximize $ \mid \sum_\sigma n_{1,\sigma} - \sum_\sigma n_{2,\sigma} \mid $. The current order parameters oscillate back and forth between the left and right panels of Fig.~\ref{fig:dimer_latt}(b). Because of the strong on-site attraction, an electron with $ \sigma=\uparrow $ is always accompanied by one with $ \sigma=\downarrow $.Thus, this mode is regarded as a pair breathing mode or a bipolaronic breathing mode. 

For even larger $ \mid U \mid $ or smaller transfer integrals, the slowly varying component becomes even slower and the breathing motion becomes more evident in the time profiles of the current order parameters (not shown). In the limit of small transfer integrals, we can estimate this frequency. Two electrons with opposite spins are tightly bound and transferred in the second order with respect to transfer integrals; thus, their effective transfer integral is given by the relation 
\begin{equation}
t^{\mbox{eff}}_{b} = 2t^2_{b}/\mid U \mid 
\;, \label{eq:t_eff}
\end{equation}
for $ b=1 $, 2, and $ y $. The factor 2 comes from the two second-order processes (an electron with spin up first or an electron with spin down first). However, in the present approximation, one-body wave functions with spin up and spin down are always degenerate, and the factor 2 is lost. In the strong-coupling limit, the difference between the lowest unoccupied and highest occupied energy levels in the mean-field ground state $ 2\Delta $ becomes $ \mid U \mid $. For large $ \mid U \mid $ but not in this limit, the estimation becomes better if the energy denominator $ \mid U \mid $ is replaced by $ 2\Delta $. Thus, in the present approximation, we use 
\begin{equation}
t^{\mbox{eff,MF}}_{b} = t^2_{b}/(2\Delta) 
\;, \label{eq:t_eff_MF}
\end{equation}
for $ b=1 $, 2, and $ y $. In Fig.~\ref{fig:c_um2p0_and_11t4500_3xxx}(b), we plot the frequency of the pair breathing mode and its estimation on the basis of $ t^{\mbox{eff,MF}}_{b} $ ($ b=1 $, 2, and $ y $) 
\begin{equation}
\omega^{\mbox{pair,MF}}_{\mbox{osc}} = 2\left( \mid t^{\mbox{eff,MF}}_1 \mid + \mid t^{\mbox{eff,MF}}_2 \mid +2 \mid t^{\mbox{eff,MF}}_y \mid \right) 
\;. \label{eq:w_pair_MF_osc}
\end{equation}
They coincide with each other in this limit. It is confirmed that when the exact diagonalization method is used for a 12-site chain in the present model with $ t_y = V_{\mbox{sp}} = V_{\mbox{ch}} = V_{\mbox{ph}} = 0 $ and periodic boundary conditions, the pair breathing mode appears after strong excitation for large $ \mid U \mid $ and its frequency is given by 
\begin{equation}
\omega^{\mbox{pair}}_{\mbox{osc}} = 2\left( \mid t^{\mbox{eff}}_1 \mid + \mid t^{\mbox{eff}}_2 \mid \right) 
\; \label{eq:w_pair_osc}
\end{equation}
in the strong-coupling limit (not shown). The relationship between the frequencies determined from the exact-diagonalization spectra and the numerical values given by Eq.~(\ref{eq:w_pair_osc}) is similar to that shown in Fig.~\ref{fig:c_um2p0_and_11t4500_3xxx}(b). The appearance of the pair breathing mode in the exact diagonalization study implies that the symmetry breaking accompanied by superconductivity is not required for its appearance. 

\subsection{Middle-frequency charge-oscillation mode}
\begin{figure}
\includegraphics[height=18.0cm]{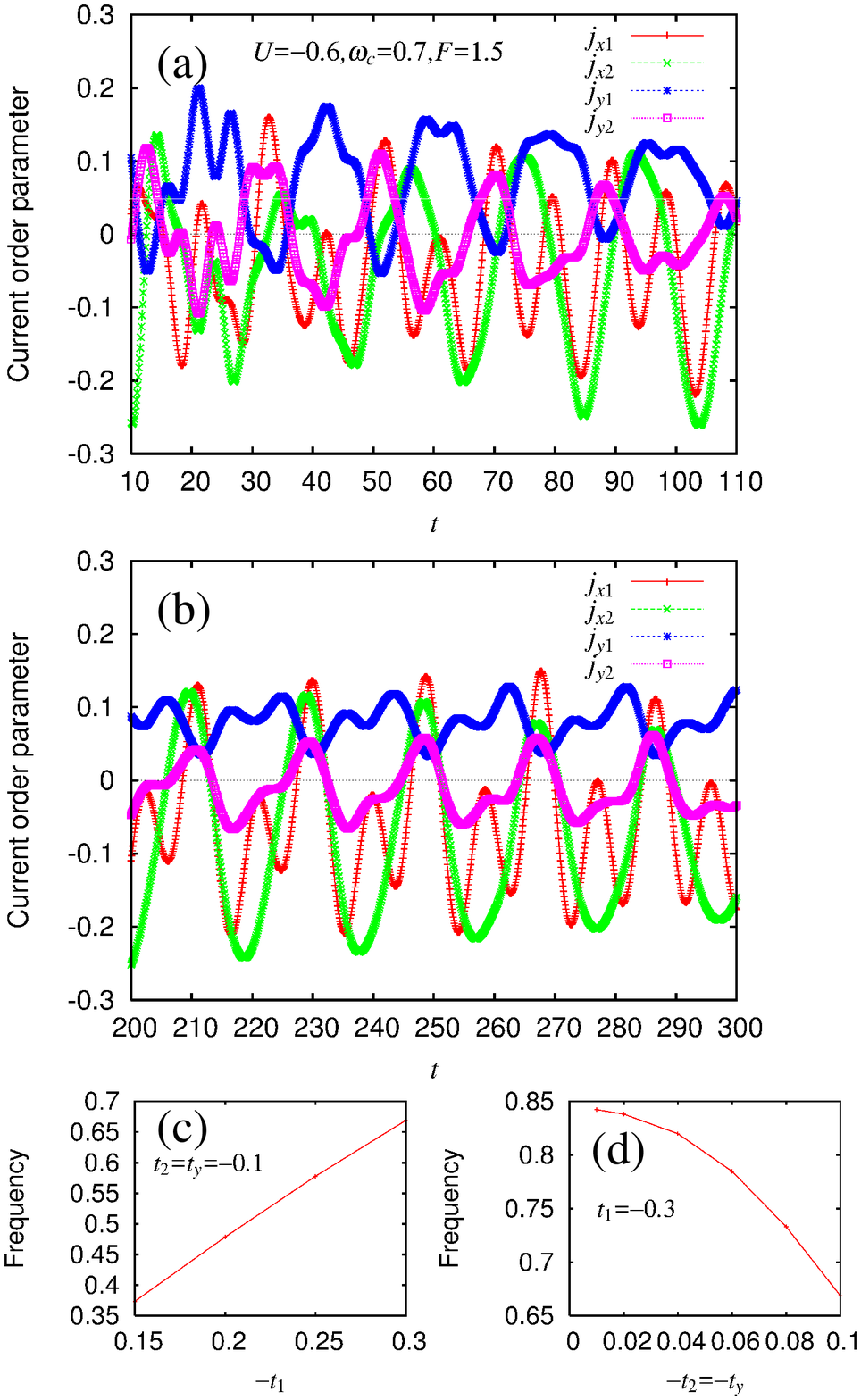}
\caption{(Color online) 
(a) Time profiles of current order parameters, $ j_{x1} $, $ j_{x2} $, $ j_{y1} $, and $ j_{y2} $, for intermediate on-site attraction $ U = -0.6 $ with $ t_1 = -0.3 $ and $ t_2 = t_y = -0.1 $, (b) Time profiles similar to (a) but for a different time domain. Frequency of middle-frequency charge-oscillation mode (c) as a function of $ \mid t_1 \mid $ with $ t_2=t_y=-0.1 $ and (d) as a function of $ \mid t_2 \mid $ with $ t_2=t_y $ and $ t_1=-0.3 $, for $ U = -0.6 $. The other parameters are $ \omega_c=0.7 $ and $ F=1.5 $. For (c) and (d), the time span $ T < t < 500T $ is used for the Fourier spectra with higher resolution. 
\label{fig:c_n3_311_um6_w7_e1p5_11t4500_yxx_um6}}
\end{figure}

For $ U=-0.6 $, where the middle-frequency charge-oscillation mode is dominant in the Fourier spectrum, the time profiles of the current order parameters are shown in Figs.~\ref{fig:c_n3_311_um6_w7_e1p5_11t4500_yxx_um6}(a) and \ref{fig:c_n3_311_um6_w7_e1p5_11t4500_yxx_um6}(b) for different time domains after photoexcitation. 
The quantity $ j_{y1} $ is often opposite in sign to the other three. This corresponds to a current distribution where the (red) arrows for $ j_{x1} $ are reversed in Fig.~\ref{fig:dimer_latt}(b) that describes the breathing modes. It does not maximize $ \mid \sum_\sigma n_{1,\sigma} - \sum_\sigma n_{2,\sigma} \mid $ but it wastes charge motion. The increase (decrease) in the local charge density caused by $ j_{x1} $ is partially canceled by $ j_{x2} $, $ j_{y1} $, and $ j_{y2} $. Sometimes the relative signs are not like this, but $ j_{x1} $ almost always has the same sign as $ j_{x2} $, and $ j_{y1} $ and $ j_{y2} $ almost always have opposite signs for any time domain after photoexcitation. 

Now, we show the dependence of the frequency of the middle-frequency charge-oscillation mode on transfer integrals. The frequency increases as the magnitude of the intradimer transfer integral $ \mid t_1 \mid $ increases [Fig.~\ref{fig:c_n3_311_um6_w7_e1p5_11t4500_yxx_um6}(c)], while it decreases as the magnitude of the interdimer transfer integral $ \mid t_2 \mid $ or $ \mid t_y \mid $ increases [although $ \mid t_2 \mid $ and $ \mid t_y \mid $ are simultaneously varied in Fig.~\ref{fig:c_n3_311_um6_w7_e1p5_11t4500_yxx_um6}(d)]. Similar dependences are also obtained for $ U=-0.4 $ and $ U=-0.7 $ (not shown). These facts are consistent with the fact that the increase (decrease) in the local charge density caused by $ j_{x1} $ ($ t_1 $ process) is partially canceled by $ j_{x2} $ ($ t_2 $ process), $ j_{y1} $, and $ j_{y2} $ ($ t_y $ processes). Note that as $ \mid t_2 \mid $ and $ \mid t_y \mid $ decrease, the frequency of the electronic breathing mode (the so-called high-frequency charge-oscillation mode) decreases and that of the present mode (the so-called middle-frequency charge-oscillation mode) increases. Consequently, the frequency of the former can be lower than that of the latter,  contrary to their naming. 

\section{Conclusions and Discussion}
Following the previous study,\cite{yonemitsu_jpsj18a} where a high-frequency charge-oscillation mode is shown to appear after an intense electric-field pulse is applied to electron systems on dimer lattices including those of organic conductors $\kappa$-(BEDT-TTF)$_2$X, we have searched for charge-oscillation modes that appear in superconductors under similar conditions. Using the Hartree-Fock-Gor'kov approximation, the time-dependent Schr\"odinger equation is numerically solved, and the time profiles of charge densities after photoexcitation are analyzed through their Fourier spectra. To characterize the charge-oscillation modes, we show transient current distributions and how their frequencies depend on model parameters. 

For weak attractions, the charge-oscillation mode has a high frequency and is regarded as an electronic breathing mode, whose frequency is already known as a function of transfer integrals for on-site interactions.\cite{yonemitsu_jpsj18a} Its amplitude becomes smaller when the pairing order parameter is increased by modifying interactions. This is due to the competition between the charge-density difference and the pairing order parameter. Although the present pairing is $ s $-wave and interaction strengths are varied here, this fact seems consistent with the fact that the corresponding photoinduced increment in reflectivity decreases as the temperature decreases below the superconducting transition temperature.\cite{kawakami_np18} 

For strong attractions, two electrons with opposite spins move together. Consequently, the charge-oscillation mode has a low frequency and is regarded as a pair breathing mode. In fact, the transfer integral $ t_{b} $ in the equation for the frequency of the electronic breathing mode is replaced by the {\em effective} transfer integral $ t^{\mbox{eff,MF}}_{b} $ for a pair of electrons in the equation for the frequency of the pair breathing mode in the strong-coupling limit. Thus, the latter frequency decreases with increasing on-site attraction. 

For intermediate strengths of attractions, another charge-oscillation mode is found to appear after an intense electric-field pulse is applied. In this case, an electron cannot move alone, but a pair of electrons is not so tightly bound. The dependence of this frequency on the model parameters is different from those of the breathing modes. It increases with the on-site attraction and the magnitude of the intradimer transfer integral, and it decreases as the magnitudes of the interdimer transfer integrals increase. This fact is consistent with the transient current distributions. 

Although the two new charge-oscillation modes are concerned with $ s $-wave pairing, we expect similar charge-oscillation modes for other pairings in addition to the electronic breathing mode if the lattice has a dimerized structure. If such a mode is experimentally observed, it would optically contribute to the characterization of superconductivity, e.g., where it is located in the BCS-BEC crossover. As discussed already,\cite{yonemitsu_jpsj18a,nag_arx} the emergence of a strong-field-induced charge oscillation is regarded as a synchronization phenomenon. This aspect will also be studied in the future. 

\begin{acknowledgments}
The author is grateful to S. Iwai and Y. Tanaka for various discussions. 
This work was supported by Grants-in-Aid for Scientific Research (C) (Grant No. 16K05459) and Scientific Research (A) (Grant No. 15H02100) from the Ministry of Education, Culture, Sports, Science and Technology of Japan. 
\end{acknowledgments}

\bibliography{68913}

\end{document}